\documentclass[slac_one]{revtex4}
\usepackage{graphicx}
\usepackage{fancyhdr}
\begin{document}
\title{Alignment of the ATLAS barrel muon spectrometer}
\author{F. Chevallier (for the ATLAS collaboration)}
\affiliation{CEA, Irfu, SPP Centre de Saclay, F-91191 Gif-sur-Yvette, FRANCE}

\begin{abstract}
The muon spectrometer of the ATLAS experiment is one of the largest detectors ever built. At the LHC, new physics signs could appear through high momentun muons (1~TeV). Identification and precise momentum measurement of such muons are two of the main challenges of the ATLAS muon spectrometer.\\
In order to get a good resolution for high energy muons (i.e. 10\% at 1 TeV), the accuracy on the alignment of precision chambers must be of the order of 50 microns. Several procedures have been developed to reach such a precision. This document describes complementary techniques used to align the muon sub-detectors, and their results : the optical system, the muon cosmic rays and the straight tracks coming from collisions.
\end{abstract}

\maketitle

\thispagestyle{fancy}

\section{INTRODUCTION}
\begin{figure}[pht]
\centering
\includegraphics[width=28mm]{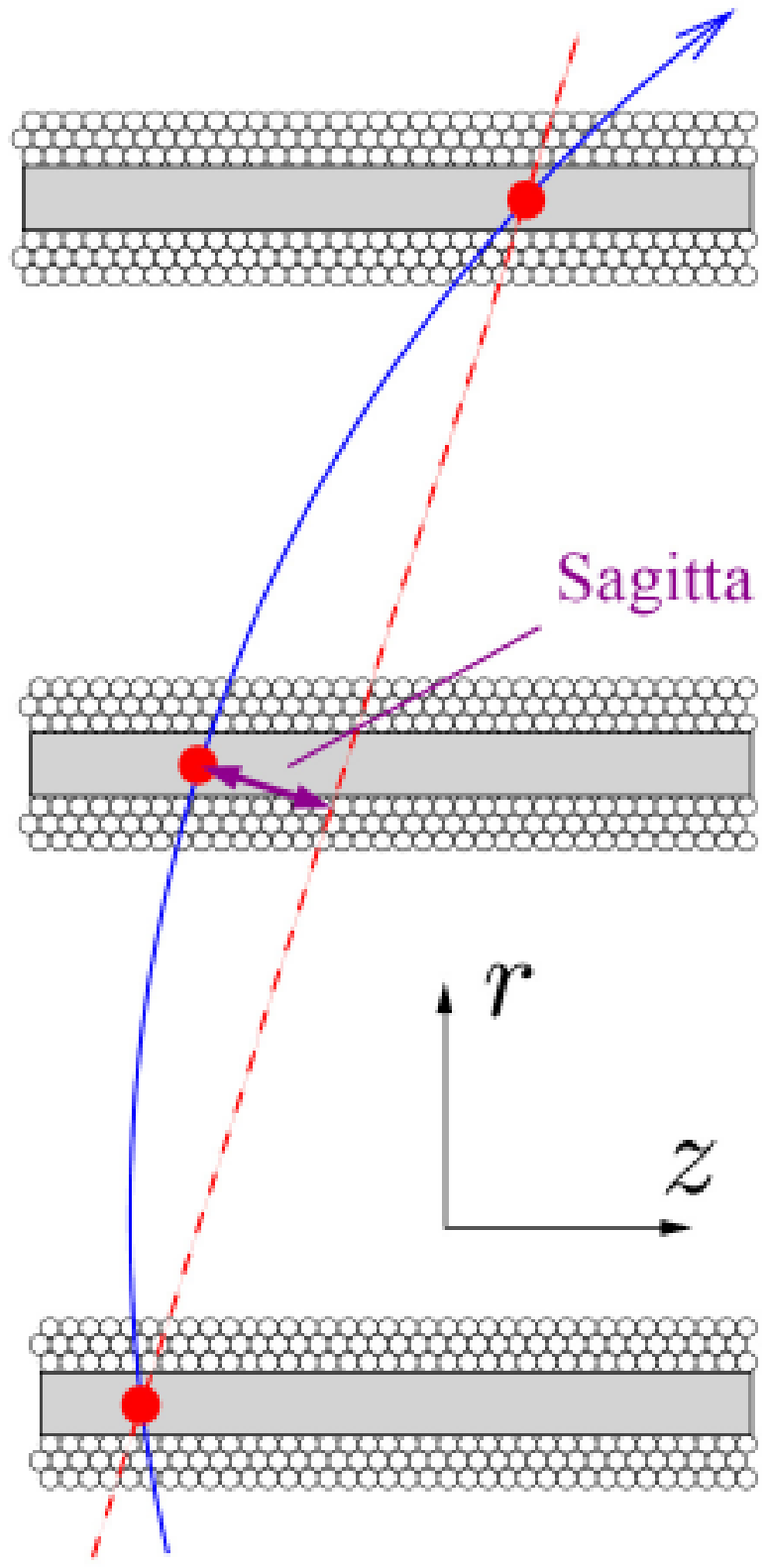}
\hspace{2cm}
\includegraphics[width=59mm]{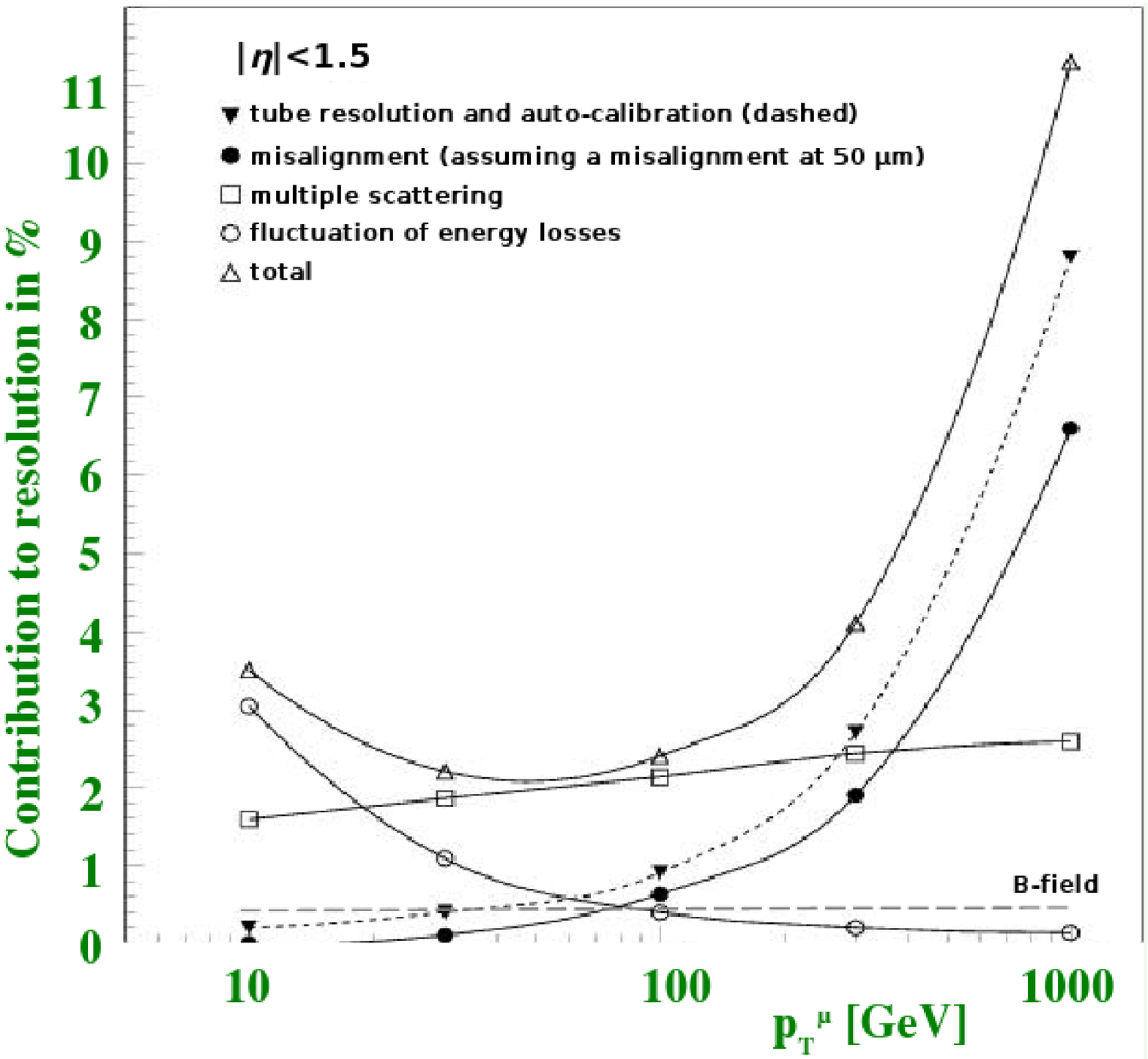}
\caption{\label{Intro} Left: Illustration of the muon transverse momentum measurement, using the three chambers of MDT. Right: The contributions to the momentum resolution as a function of transverse momentum, assuming a misalignment of $\rm 50~\mu m$~\cite{ref:MuonTDR}.}
\end{figure}

The muon spectrometer measures the deflexion of the muon tracks, bent by a toroidal magnetic field. Then, the transverse momentum of the muon is computed from the measured curvature of the track. The high precision tracking chambers are arranged in such a way that particles coming from the interaction point cross three layers of chambers (Inner, Middle, Outer) as shown in fig~\ref{Intro}. These chambers use Monitored Drift Tubes ({\it MDT}) technologies.

The muon spectrometer has been designed to satisfy physics requirements, and background and environmental conditions. In particular, the resolution on the measured momentum of 1~TeV muons is $\rm \delta p_T/p_T=10\%$. According to fig~\ref{Intro}, the contribution to the momentum resolution from the chamber misalignment becomes comparable to the other intrinsic sources if the misalignment between chambers is of the order of $\rm 50~\mu m$. In other words, as the sagitta of 1~TeV muons is around $\rm 500~\mu m$, the sagitta uncertainty must be smaller than $\rm 50~\mu m$. In order to fulfil these tight requirements, three complementary techniques are used in ATLAS. They are explained in this document.
\vspace{-0.7cm}


\section{ALIGNMENT WITH OPTICAL LINES}
In order to reach a precision of $\rm 50~\mu m$ on chamber alignment, the distances and orientations between chambers are measured with a set of optical sensors. This system is decribed in this section as well as its goals.\\

\begin{figure*}[pht]
\includegraphics[width=90mm]{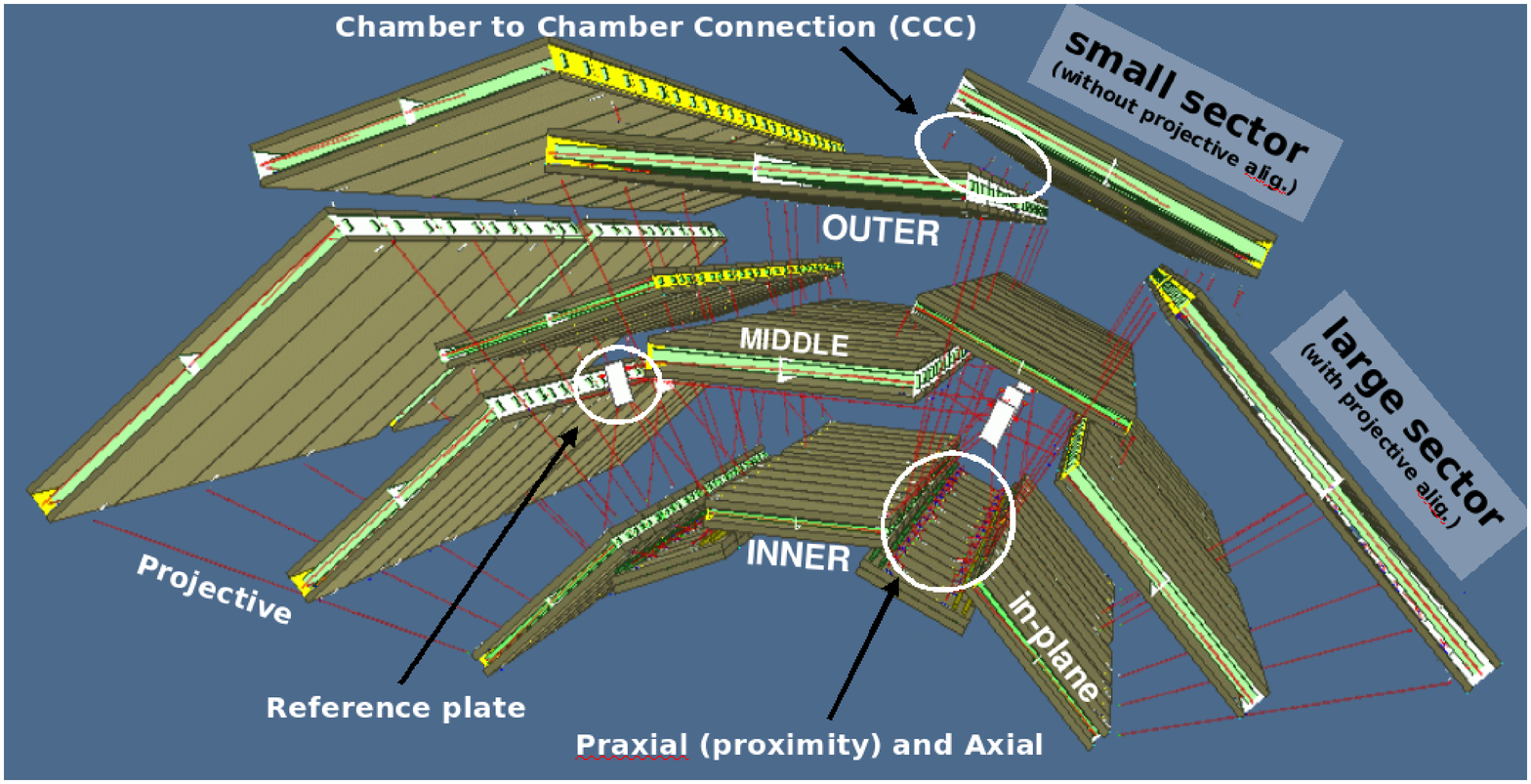}
\includegraphics[width=75mm]{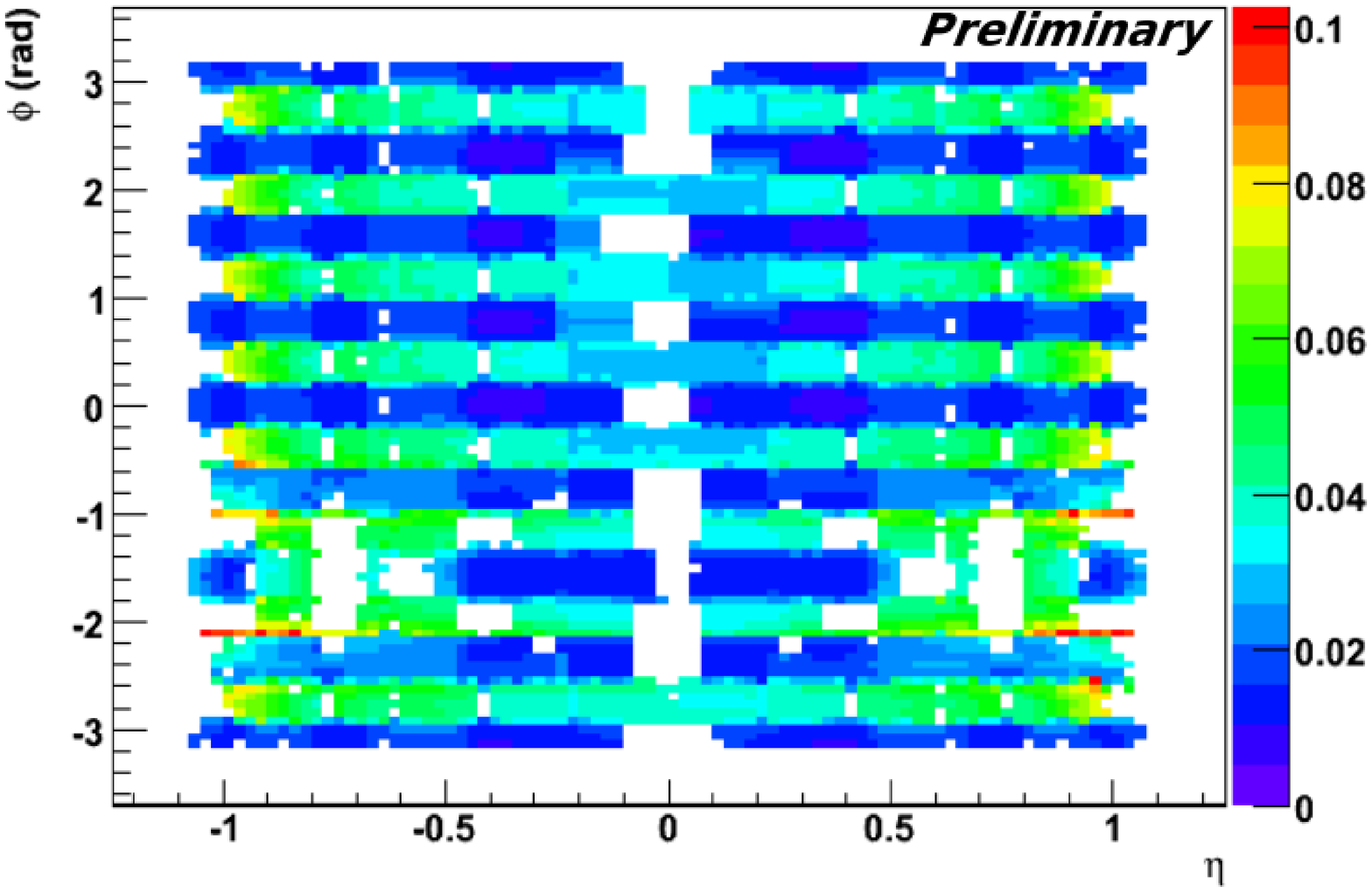}
\caption{\label{OpticalSystem} Left: 3D view of the upper sectors in the barrel spectrometer and the optical system. Right: Error on sagitta (in mm) coming from the uncertainties on the chamber positions, assuming a perfect absolute calibration of the optical systems~\cite{ref:PFG}.}
\end{figure*}

The optical system connects the chambers to their neighbors with optical lines, as shown in fig.~\ref{OpticalSystem}. In addition to these lines, the deformations of chambers is measured by an internal optical system. These measurements are based on a 3-point system (2D-coded mask or spot + lens + video sensor). They give the translations of the neighboring chambers perpendicular to the optical axis, the rotation angle around the optical axis and the distance between chambers. In total, about 6,000 optical lines are installed in the barrel spectrometer.

In order to obtain the real geometry of the whole spectrometer, all the constraints from the optical systems are transformed in geometric relations between chambers. Additional measurements give important external constraints (the survey system) or internal constraints (in-situ mechanical measurements). Finally, the positions of the 622 barrel chambers are obtained using a $\chi^2$ minimization routine. The fig.~\ref{OpticalSystem} shows the results of this work.

The optical system has been developed to precisely measure the geometry of the muon spectrometer and to follow its deformations with time. In order to have an absolute alignment, the first step consists in obtaining the initial geometry. The initial chamber positioning is accurate to roughly 5~mm and 2~mrad~\cite{ref:PFG}. These uncertainties on the alignment are two orders of magnitude above the $\rm 50~\mu m$ goal because of the following issues: low accuracy of a few sensor position, calibration lost during the installation, broken sensors, inversions in the cabling system, dust on masks. That is why controls, cross-checks and combination with other systems (see next sections) are needed to reach $\rm 50~\mu m$.

\noindent Once the initial geometry is obtained, the relative displacements of chambers will be measured to give the geometry at any time. The optical system is able to follow displacements with a precision of $O\rm (10~\mu m)$ and $O(10~\mu rad)$. These geometry changes are caused by the magnetic field (on/off), temperature variations, ...\\

The installation of the optical elements in the cavern is almost finished and the debugging of the alignment of a few octants has already started. In the first fully debugged sector of the barrel, the present understanding of the absolute positioning is below $\rm 300~\mu m$.
\vspace{-0.5cm}

\section{ALIGNMENT WITH COSMIC TRACKS}
The atmospheric (cosmic) muons are helpful for the alignment of the muon spectrometer because they produce straight tracks when the toroidal magnetic field is off. Neglecting the effect of multiple scattering, the sagitta of these muons must be zero.

The idea for the alignment is to compute an estimator that becomes minimum when the true position of chambers is tested. The procedure is based on the quality of the reconstructed cosmic tracks, between the geometry described in the database and the true one. This latter is found either by $\chi^2$ minimization or by the {\sc MILLEPEDE}~\cite{ref:IP} method.\\

The alignment of the muon spectrometer faces several challenges and physics issues. Firstly, this analysis with ATLAS real data requires a good understanding of the detector and reconstruction effects. It is the first test of a full ATLAS sub-detector in operating conditions. Secondly, it is not possible to align the full spectrometer with the cosmics, because the cosmic muon flux is too low in some chambers. Finally, the material in the muon spectrometer induces a non-negligible multiple scattering for the cosmic muons. The sagitta coming from multiple scattering is greater than 10 mm for half of these cosmics muons. So, the multiple scattering is the limiting effect for the alignment of the muon spectrometer with cosmic tracks. It is preferable to use muons with higher momentum (see next section).\\

To conclude, the ATLAS cosmics runs have improved our understanding of the detector effects. The alignment has been tested with 10,000 cosmic tracks per chamber. The average sagitta error on cosmic tracks is $O(50~\mu m)$ with this alignment, but the multiple scattering leads to larger sagitta errors. However, the comparison between present mechanical measurements and this alignment is successful at $O(100~\mu m)$.
\vspace{-0.5cm}

\section{ALIGNMENT WITH STRAIGHT TRACKS FROM COLLISIONS}
\begin{figure*}[pht]
\vspace{-0.5cm}
\includegraphics[width=60mm]{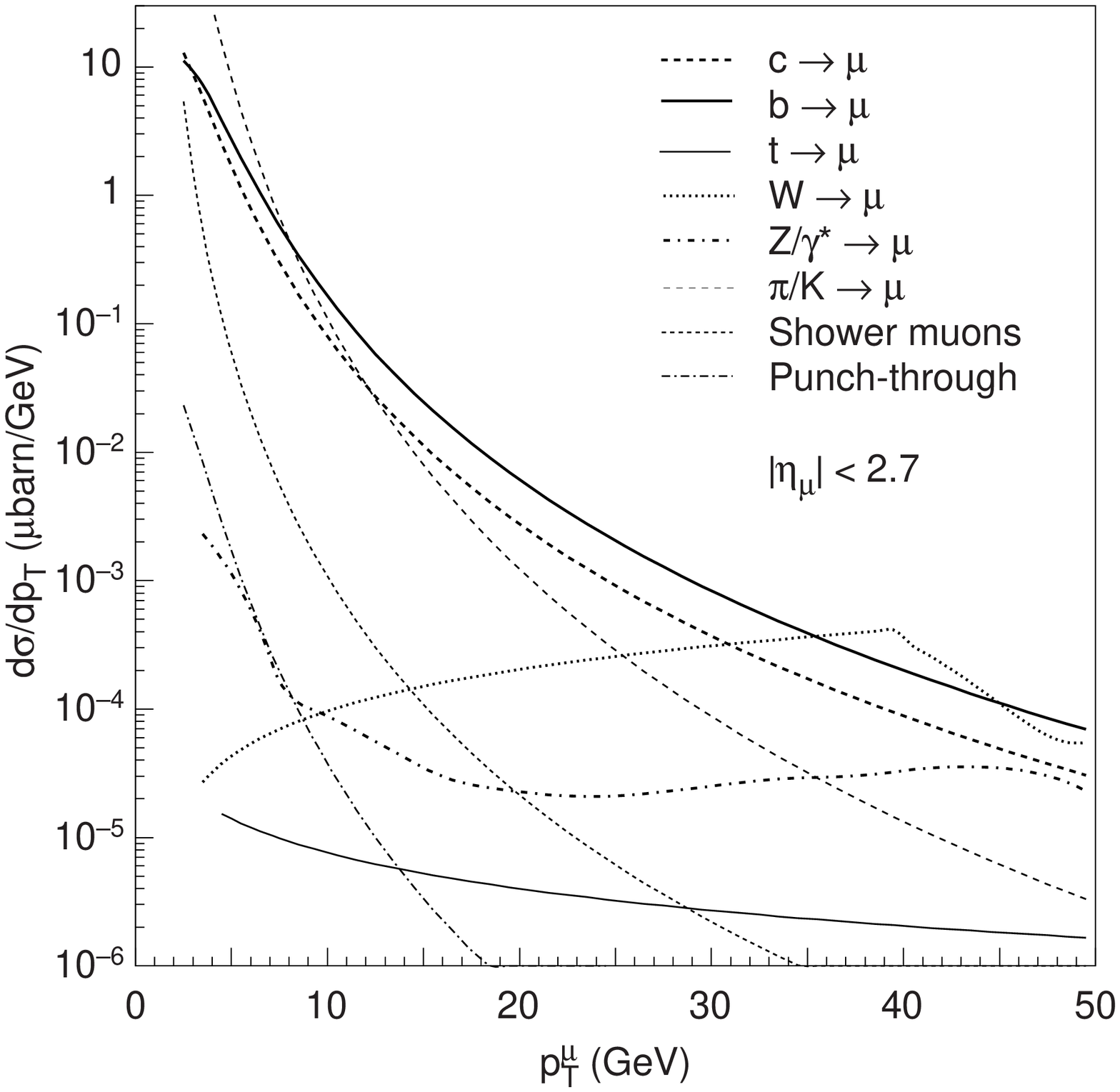}
\hspace{1cm}
\includegraphics[width=63mm]{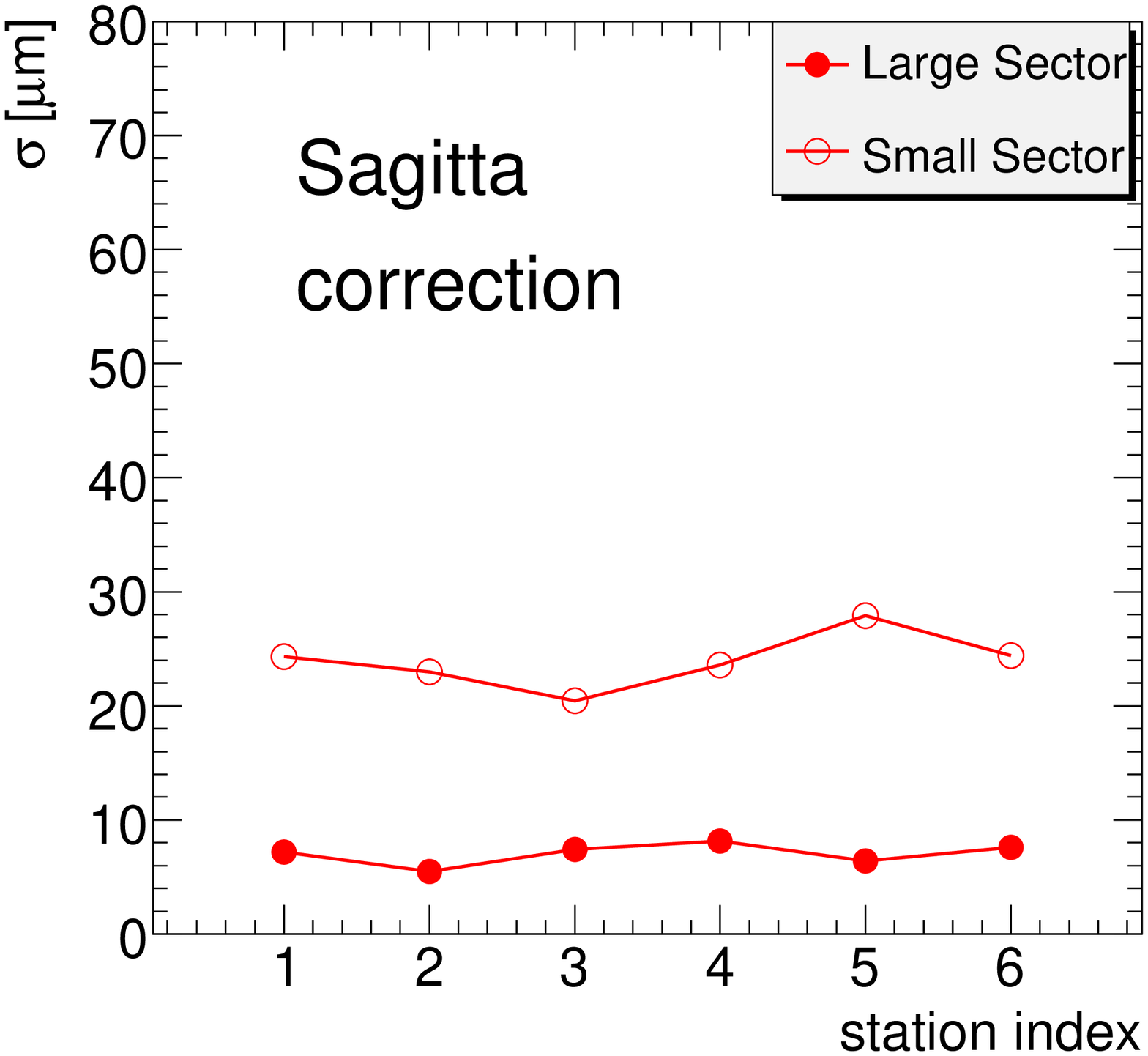}
\caption{\label{LHCData} Left: Transverse momentum dependence of inclusive muon cross-sections~\cite{ref:MuonTDR}. Right: Statistical uncertainty on sagitta correction using 20 GeV projecting tracks in six stations of the spectrometer.~\cite{ref:IP}}
\end{figure*}

The alignment with straight tracks, from special runs with the toroid magnetic field switched off, is required to obtain the initial geometry. The collisions at the LHC will produce high energy muons, as shown in fig.~\ref{LHCData}, and the effect of multiple scattering will be smaller than with cosmics. These straight tracks from collision data are aimed at complementing the optical measurements.

In order to test the alignment procedure with straight projective tracks, single muons with $\rm p_T=3~and~20~GeV$ have been generated and fully simulated. The {\sc MILLEPEDE} algorithm has been used to assess the performance of the alignment. The results of this study are shown fig.~\ref{LHCData}. The statistical uncertainty on the sagitta correction is smaller than $\rm 30~\mu m$ for 20 GeV tracks. For small sectors, they are a factor five larger in comparison with large sectors. The main systematic uncertainties come from the deformation of chambers, the temperature variations and the wire sagging inside the MDT. The effect of the multiple scattering is averaged by the large statistics.\\

To conclude, the simulated data has shown that reaching an accuracy of $\rm 30~\mu m$ on the sagitta should be possible using 100,000 projective tracks of 20 GeV muons per sector and the optical constraints.
\vspace{-0.5cm}

\section{CONCLUSION}
The ATLAS muon alignment community is working on the optical system in order to get the design precision of $\rm 50~\mu m$ on the muon sagitta. For the initial knowledge of the spectrometer geometry, the cosmic muons and straight track data are mandatory to reach this precision.
\vspace{-0.5cm}

\section{ACKNOWLEDGMENTS}
I would like to thank the members of the ATLAS muon alignment community, and in particular my colleagues at IRFU-Saclay and NIKHEF institutes. Many thanks to A. Marzin who presented the poster at the ICHEP conference.

\end{document}